\documentclass[superscriptaddress,reprint,citeautoscript,prb]{revtex4-1}
\usepackage{graphicx} 
\usepackage[margin=1in]{geometry} 
\usepackage{amsmath}
\usepackage{hyperref} 
\usepackage{multirow}
\usepackage{newtxtext}                                        
\usepackage{booktabs}
\usepackage{xcolor} 
\usepackage{hyperref}
\usepackage{mathtools}
\usepackage[normalem]{ulem}
\usepackage{comment}
\hypersetup{colorlinks, 
	linkcolor={blue!75!black!80!yellow},
	citecolor={blue!75!black!80!yellow}, 
	urlcolor={blue!75!black!80!yellow}
	}
\usepackage[cmintegrals]{newtxmath}
\makeatletter \renewcommand\@make@capt@title[2]{\@ifx@empty\float@link{\@firstofone}{\expandafter\href\expandafter{\float@link}}\sffamily{\textbf{#1}}\@caption@fignum@sep#2 }

\newcommand{\HarvardSEAS}{John A. Paulson School of Engineering and Applied
Sciences, Harvard University, Cambridge, MA, USA}
\newcommand{\HarvardCCB}{Department of Chemistry and Chemical Biology, Harvard
University, Cambridge, MA, USA} 
\newcommand{\MIT}{Department of Electrical Engineering and Computer Science,
Massachusetts Institute of Technology, Cambridge, MA, USA}
\newcommand{\cvvv}{C\textsubscript{3v}}
\newcommand{\chh}{C\textsubscript{2h}}
\newcommand{\dddd}{D\textsubscript{3d}}
\newcommand{\Vm}{V\textsuperscript{--}}
\newcommand{\Vn}{V\textsuperscript{0}}

\usepackage[textwidth=1.5cm,textsize=footnotesize]{todonotes}

\begin{document} 

\author{Isaac Harris}
\thanks{These authors contributed equally}
\affiliation{\HarvardSEAS}\affiliation{\MIT}
\author{Christopher J. Ciccarino}
\thanks{These authors contributed equally}
\affiliation{\HarvardSEAS}\affiliation{\HarvardCCB}
\author{Johannes Flick}\affiliation{\HarvardSEAS}
\author{Dirk R. Englund}\affiliation{\MIT}
\author{Prineha Narang}\email{prineha@seas.harvard.edu}\affiliation{\HarvardSEAS}

\title{Group III Quantum Defects in Diamond are Stable Spin-1 Color Centers}

\date{\today}

\begin{abstract} 

Color centers in diamond have emerged as leading solid-state 
``artificial atoms'' for a range of quantum technologies, from quantum
sensing to quantum networks. Concerted research activities 
are now underway to identify new color centers that
combine stable spin and optical properties of the
nitrogen vacancy (NV\textsuperscript{--}) with the spectral stability of
the silicon vacancy (SiV\textsuperscript{--}) centers in diamond, with recent research identifying other group IV color centers with superior properties.
In this \emph{Letter}, we investigate a new class of diamond quantum emitters from first principles, the group III color centers, which we show to be thermodynamically stable in a spin-1, electric-field insensitive structure.
From \emph{ab initio} electronic structure methods, we characterize the product Jahn-Teller (pJT) effect present in the excited state manifold of these group III color centers, where we capture symmetry-breaking distortions associated with strong electron-phonon coupling. 
These predictions can guide experimental identification of group III vacancy centers and their use in applications in quantum information science and technology.
\end{abstract}

\maketitle
Diamond color centers are promising building blocks for applications such as quantum sensing and quantum communication,\cite{Jelezko2006,Weber2010,Aharonovich:2016yu,Awschalom2018} such as in the recent demonstration of the generation of entanglement between two distant diamond nitrogen vacancy (N\Vm) centers at a rate faster than their decoherence\cite{humphreys2018}. 
While this result marks a major milestone towards the development of scalable quantum networks,
the entanglement generation rate could be improved by orders of magnitude with a high-efficiency photonic
interface to the emitter's coherent optical transition. 
One route is to couple the N\Vm center to a micro- or nano-cavity to enhance spontaneous emission into the
coherent zero phonon line, but more work is needed to overcome the N\Vm's spectral diffusion near
surfaces~\cite{atature2018}, which arises primarily from its \cvvv\ symmetry, giving 
rise to a permanent electric dipole moment. 

Another approach, which is considered here, is to develop alternative color centers that have stable optical and spin properties.
 Within the diamond material system, the group IV vacancy centers (Si\Vm, Ge\Vm, Sn\Vm, and Pb\Vm) have been characterized 
experimentally~\cite{Neu:2013,Inubushi2015,Siyushev2017,Trusheim2019} and theoretically~\cite{Thiering2018a,gali:SiV:2013,Thiering2019}.
The group IV negative centers adopt an inversion-symmetric split-vacancy structure which has no permanent electric dipole moment,
making their optical transitions less sensitive to electric field noise near surfaces~\cite{Sipahigil2014}.
Additionally, the branching ratio into the zero phonon line (ZPL) can be more than an order of magnitude
higher than for the NV center, though this gain is partially offset by a lower internal quantum efficiency. 
Unfortunately, the negatively charged group IV centers suffer from a 
phonon-mediated dephasing mechanism of the ground state spin structure, which limits their spin coherence times~\cite{Doherty2015}.
Recent findings suggest that by careful boron doping of diamond, the SiV can favor a neutral charge with the stable spin-1 electronic
ground state of the N\Vm\ center and the
stable optical transitions of the Si\Vm~\cite{rose:SiV0:2018}, though more work is needed to demonstrate both properties in the same emitter.
A natural question is whether there are other color centers that exhibit stable optical and spin properties in their thermodynamically
favored charge state in intrinsic diamond. Outside of the group IV centers, previous work has characterized other
dopant-vacancy centers in diamond~\cite{Goss2005};
however their optoelectronic properties are not yet understood.

In this \textit{Letter}, we report predictions of the diamond group III vacancy defects XV, with X = Al, Ga, In, and Tl, and characterize their properties using \emph{ab initio} electronic structure theory of the ground and excited 
state manifolds. Our calculations reveal that the three lightest defects are stable in the high-symmetry \dddd\ configuration, with 
a thermodynamically preferred --1 charge state which makes them isoelectronic with the Si\Vn\ defect.
We characterize the product Jahn-Teller (pJT) effect~\cite{Qiu2001, Thiering2019} present in the excited state
manifold, where we capture symmetry-breaking distortions associated
with strong electron-phonon coupling. We discuss the impact of these distortions,
in particular on the predicted zero-phonon line energies. We also capture vibronic spectra and discuss the relatively
high predicted ZPL emission efficiency found. Overall, the group III vacancy centers are found to be similar to their group IV neutral counterparts, but with a spin-1 ground state for the thermodynamically stable negative charge state in intrinsic diamond. The combination of the stable spin-1 ground state and symmetry-protected optical transitions makes these centers excellent candidates for quantum technologies. 

\begin{figure*}
\includegraphics[width=\textwidth]{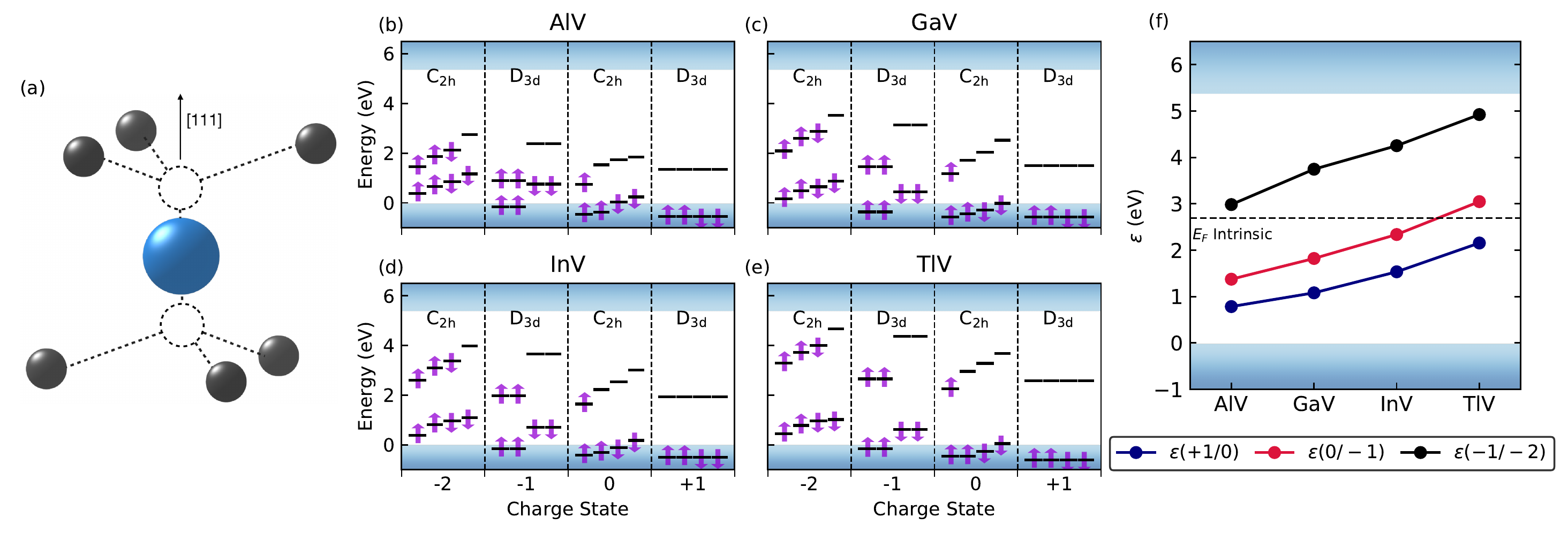}
\caption{\textbf{(a)} \dddd\ ground state structure of a group III vacancy defect in diamond. The impurity atom lies 
directly between two carbon empty lattice sites, equidistant to six nearest neighbor carbons.
\textbf{(b-e)} Ground state spin resolved energy level structure of the group III vacancies in charge states -2 to +1. Both the --1 and +1 charge states are stable in the 
high symmetry \dddd\ structure, while the --2 and neutral charge states have uneven occupations and distort to a \chh\ structure. {\bf (f)}
Predicted thermodynamic charge transition levels for
each of the group III defect centers. AlV, GaV, and InV are found to be stable at the intrinsic Fermi level in a --1 charge state, while TlV is likely to exist in either the --1 or neutral charge state.}
\label{fig:ground}
\end{figure*}

First, we investigate the electronic properties of the group III vacancies for aluminum, gallium, indium, and thallium 
using density functional theory with the HSE06 hybrid functional~\cite{QE1,QE2,ONCV,Heyd:2003,Heyd:2006}.
All four centers are found to be stable in the split-vacancy configuration shown in Fig.~\ref{fig:ground}(a). Depending on the charge state, this configuration has either a \dddd\ geometry with the principal axis oriented along the [111] crystallographic direction, or a \chh\ geometry.
Both configurations are inversion symmetric about the substituent atom, and therefore cannot have a net dipole moment, making their optical transitions insensitive to charge noise.
Each of the presented calculations uses a 512 atom `supercell' and samples the $\Gamma$ point;
we refer to the SI for computational details and convergence studies. 

As previous work~\cite{Hepp2014a,gali:SiV:2013,Thiering2018a,Thiering2019} has established for group IV emitters in diamond,
the orbitals associated with the group III defects can be identified by their symmetry. Two pairs of degenerate Kohn-Sham (KS) orbitals are localized within the diamond bandgap, having $E_u$ and $E_g$ symmetry, respectively. The orbitals within each degenerate pair are denoted as $e_{ux(y)}$ and $e_{gx(y)}$, where 
$x$ or $y$ refers to their orientation. While additional localized defect orbitals are present, they are energetically
far from the band edges and are therefore not accessible for optical excitation.

Figure~\ref{fig:ground} shows the electronic ground state configurations for these defects in charge states from --2 to +1. Panels (b-e) plot the energies of the spin majority (spin up) and spin minority (spin down) channels for each 
of the defects, where the $e_u$ orbitals lie below the $e_g$ orbitals in each case.

Of particular interest are the --1 charge states, which 
are spin-1 states, isoelectronic to the group IV neutrals~\cite{Thiering2019} and stable in a \dddd\ configuration.
For each defect, the ground state electronic configuration of the --1 charge state is denoted by $e_u^4 e_g^2$, such that 
there are two unfilled $e_g$ orbitals, both of the same spin channel. 
This can equivalently be formulated in terms of a pair of holes occupying the remaining $e_g$ ground state orbitals. 
In contrast to \emph{ab initio} orbital energetics of group IV neutral systems~\cite{Thiering2018a,Thiering2019}, where the $e_u$ KS orbitals
lie below the valence band edge, both the $e_u$ and $e_g$
levels for the spin-minority channel are located within the band gap for the group III negatives.
This difference could lead to improved excitation efficiency for the group III vacancy centers, as these defect-defect transitions 
are energetically isolated from any bulk states.

Next, we calculate which charge state is thermodynamically most favorable at a given doping level for the diamond host.
The Fermi level at which a transition becomes thermodynamically stable is given by the charge transition levels:~\cite{freysoldt:review:2014}
\begin{equation}
\varepsilon(q_1/q_2)=
\frac{ E_{q_1}^\mathrm{tot}+E_{q_1}^\mathrm{corr}-E_{q_2}^\mathrm{corr}-E_{q_2}^\mathrm{tot} }
{q_2-q_1},
\label{eq:CTL}
\end{equation}
with $E_{q}^\mathrm{tot}$ being the total energy from the supercell calculation in charge state $q$, and $E_{q}^\mathrm{corr}$ the corresponding charge correction to
account for periodic interaction of charges between neighboring supercells~\cite{sundararaman_first-principles_2017,freysoldt_fully_2009}.
We apply the correction techniques described in Ref.~\citenum{sundararaman_first-principles_2017}, yielding the results summarized in Fig.~\ref{fig:ground}(f). 

For aluminum, gallium, and indium substituents, the intrinsic Fermi level lies between the transition energies $\varepsilon(0/-1)$ 
and $\varepsilon(-1/-2)$, indicating that the --1 charge state is thermodynamically stable in intrinsic diamond. Following a similar trend to the group 
IV elements~\cite{Thiering2018a}, the transition energies are higher for the heavier elements, resulting in the neutral charge state being favored for thallium. This 
neutral charge state is spin \textonehalf\ and relaxes into a \chh\ symmetry due to its uneven orbital occupation (i.e. it is Jahn-Teller unstable). The 
moderate nitrogen doping typically present in diamond samples means that we might still expect to see the desirable --1 charge state in as-grown diamond for TlV, since
the transition lies close to the intrinsic Fermi level.

Given that these color centers are isoelectronic with the
Si\Vn, the negatively charged group III vacancy centers are expected to have similar electronic properties to the 
Si\Vn\ in the ground and excited states. 
In particular, exciting an electron from the lower-energy $e_u$ to the 
higher energy $e_g$ orbital means that the electronic configuration becomes
$e_{u}^3e_{g}^3$, producing an unequal occupation of the two degenerate orbital pairs. Such a system is expected to be Jahn-Teller 
unstable~\cite{bersuker:book:1990} as a result of a product of two the Jahn-Teller instabilities for
both the $e_u$ and $e_g$ orbitals, denoted as a
$(e_g \otimes e_u) \otimes E_g$ product Jahn-Teller effect~\cite{Thiering2019}. 
Here $E_g$ is the irreducible representation of the coupled phonon modes that produce the distortion.
Each of these modes is two-fold degenerate, and the Jahn-Teller distortion therefore resides in a two-dimensional
coordinate space that contains the high-symmetry \dddd\ point, as well as three equivalent energy minima in the \chh\ symmetry group.
Accurately capturing the Jahn-Teller effect is important in predicting the experimentally relevant zero phonon line energy associated
with optical emission, as the pJT symmetry reduction can result in a large energy change in the excited state, 
as seen in the case of the Si\Vn~\cite{Green2018,rose:SiV0:2018,Thiering2019}. 

\begin{figure}
\includegraphics[width=\columnwidth]{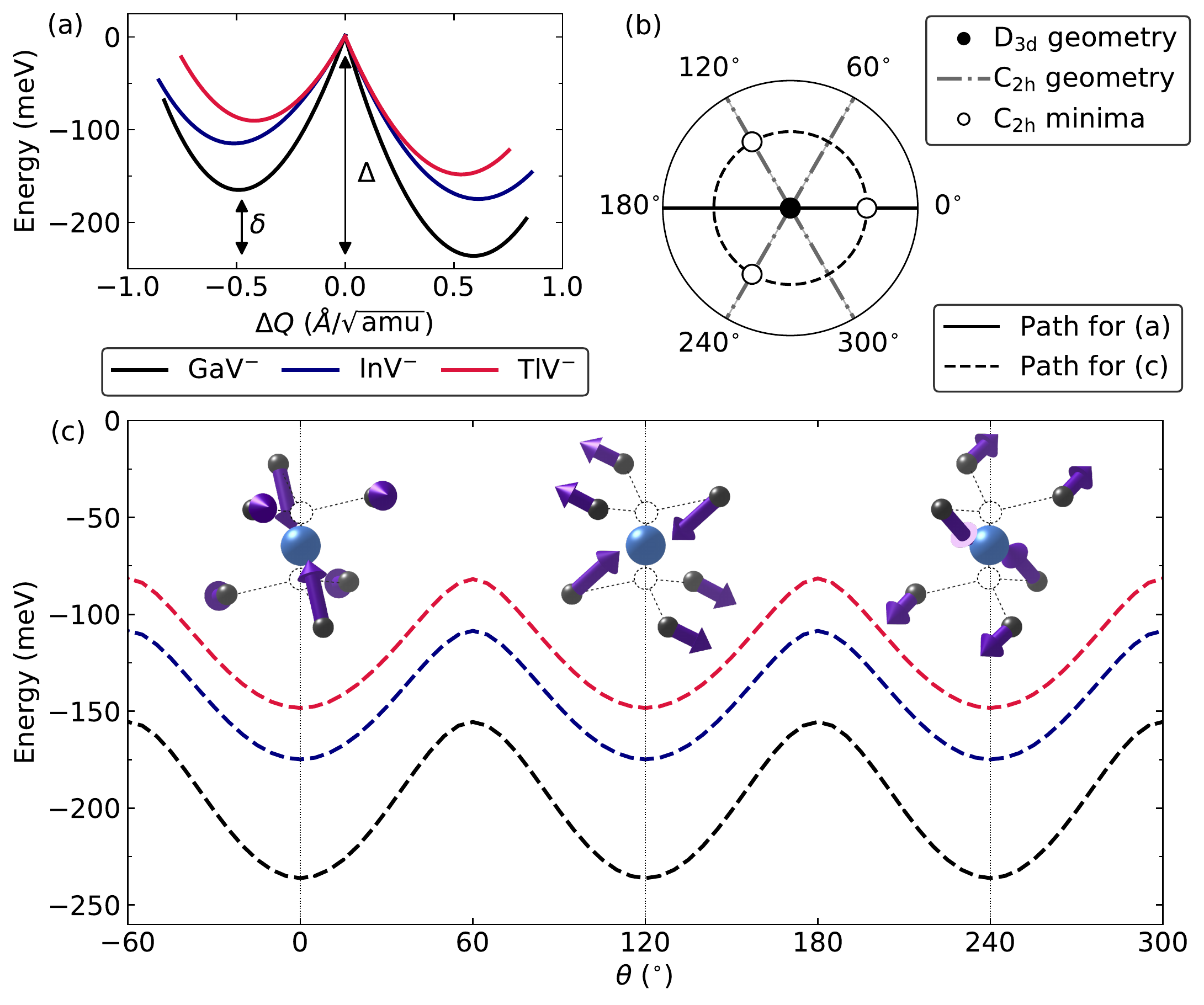}
\caption{Product Jahn-Teller instability for group III vacancy centers Ga\Vm, In\Vm\ and Tl\Vm.
\textbf{(a)} Linear interpolation between the relaxed low symmetry \chh\ and high symmetry \dddd\ geometries
for the excited state electronic configuration $e_{u}^3e_{g}^3$. The distortion is favored due to a product Jahn-Teller effect caused by the occupational imbalances in the $e_u$ and $e_g$ orbitals.
The computed instability energies are given as $\Delta$ and listed in Table~\ref{tab:pjt},
where we also list the energy barrier $\delta$ caused by higher-order phonon coupling.
\textbf{(b)} The configuration space plotted in panels (a,c) are depicted relative to the 
high symmetry \dddd\ (black dot) and three low symmetry \chh\ points (white dots).
The solid [dashed] black line indicates the path taken along the x axis for panel (a)[c].
The grey dot-dashed lines indicate the areas where the defect has \chh\ symmetry, with all points not on the lines (and not at the center \dddd\ point) 
having a C\textsubscript{i} symmetry.
\textbf{(c)} Calculated energies for a path 
equidistant to the high symmetry \dddd\ point shown in panel (b), where we find
three energetically equivalent \chh\ geometries. The local minima represent
saddle points, corresponding to the local minima shown in panel (a). 
Additionally pictured
are the three equivalent distortions from \dddd\ to 
each of the \chh\ minima.
}
\label{fig:pJT}
\end{figure}

To capture this effect in group III vacancy centers, we relax the Ga\Vm, In\Vm, and Tl\Vm\ defects
in the excited electronic state using constrained DFT ($\Delta$SCF) within Quantum Espresso~\cite{QE1,QE2}
with norm-conserving pseudopotentials~\cite{schlipf_optimization_2015} in a plane-wave basis with energy cutoff of 80 Ry. 
First, we relax the system constrained to both a \dddd\ and a \chh\ geometry with excited electronic occupations
until forces on all atoms are below $10^{-6}$ Ry/Bohr.
We additionally verify that the resulting \chh\ structure is 
the true minimum of the excited state manifold by relaxing without any symmetry constraints.
From these relaxed \chh\ and \dddd\ geometries, we can construct the full
potential energy surface associated with the Jahn-Teller modes in the excited manifold.
While the Al\Vm\ is not explicitly discussed next, it is expected to follow the same trends as the heavier group III elements.

Similar to the behavior of the group IV neutral excited states in Ref.~\citenum{Thiering2019},
there may exist four distinct excited states depending on the occupations of the defect orbitals,
as can be seen from the fact that there are two holes that can occupy 
two orbitals. However, constrained DFT is only able to access the lowest energy for a given constrained configuration,
and we therefore consider only the lowest-energy configurations of the excited state manifold
for a particular geometry.
Future extensions of this work could use a formalism along the lines of Refs.~\citenum{Kaduk2012,Plaisance2017} which restrict the symmetry of the orbitals, allowing for a better approximation of the many-body Hamiltonian and a slightly more accurate ZPL energy calculation.
Nevertheless, the information at hand allows us to calculate ZPL energies, optical spectra, and Debye-Waller factors while neglecting the relatively small many-body effects. The calculated ZPL energies, given as the energy difference between the the \chh\ excited electronic state and the \dddd\ ground electronic state,
are summarized in Table~\ref{tab:pjt}.

Figure~\ref{fig:pJT} and Table~\ref{tab:pjt} summarize the results of the pJT calculations. Panel (a) shows a one-dimensional interpolation of the excited-state potential energy surface between the \dddd\ (located at $Q=0$) and \chh\ geometries.
The global minimum is the relaxed \chh\ geometry, while the local minimum on the opposite side of the \dddd\ point is a saddle point.
The energy differences between the \dddd\ and global \chh\ minima ($\Delta$) and the saddle point and global \chh\ minima ($\delta$) are presented in Table~\ref{tab:pjt}.
In the first-order Jahn-Teller effect, the potential energy surface would be symmetrical with respect to the \dddd\ point, such that $\delta \rightarrow 0$.
However, given the relatively large values of $\delta$, higher order effects are clearly important in this system~\cite{bersuker:book:1990}.

In Fig.~\ref{fig:pJT}(c), we map out the energy of the system along a circle through the Jahn-Teller coordinate space which traverses the three energetically equivalent \chh\ minimum geometries, as well as the energy barriers between them caused by the higher-order Jahn-Teller coupling.
The trajectories in the coordinate space shown in Fig.~\ref{fig:pJT}(a,c) are visualized in Fig.~\ref{fig:pJT}(b),
where we depict the locations of high-symmetry \dddd\ geometry along with the three equivalent \chh\ configurations within the JT space.
The movement of the nearest neighbor carbons resulting from each of the equivalent distortions are shown as well.
We note that each distortion affects one pair of the nearby carbon atoms more than the other two. 
By adding each of these distortions together, we arrive at the high symmetry \dddd\ geometry once again.

\begin{table}[]
\begin{tabular}{lccc}
\hline
               & $\Delta$ (meV) & $\delta$ (meV)   & $E_{\mathrm{ZPL}}$ (nm, eV)\\ \hline
Ga\Vm          & 236              & 71             & 679, 1.82     \\
In\Vm          & 175              & 60             & 584, 2.12    \\
Tl\Vm          & 148              & 58             & 437, 2.84
\end{tabular}
\caption{Energetics of group III color centers Ga\Vm, In\Vm, Tl\Vm. The 
predicted Jahn-Teller instabilities, depicted in Fig.~\ref{fig:pJT}, are given by $\Delta$ and $\delta$. The 
zero phonon line energies are given by $E_{\mathrm{ZPL}}$ (see Fig.~\ref{fig:optical}).}
\label{tab:pjt}
\end{table}

To characterize the optical properties, we calculate \emph{ab initio} emission lineshapes using 
the method outlined in Ref.~\citenum{Alkauskas2014}. 
Phonon properties for the ground state geometries of each of the --1 charge state defects were evaluated
using a computationally more efficient PBE functional~\cite{USPP,PBE,Phonopy}, which has previously been found to be sufficiently accurate compared to the hybrid functionals used earlier~\cite{Hummer2009}.
The emission lineshape can then be calculated from the overlap of the ionic vibrational wavefunctions between the relaxed excited electronic state and the electronic ground vibrational states~\cite{alkauskas_first-principles_2012,Alkauskas2014} according to
\begin{equation}
A(\hbar\omega)=\sum_m\left|\left\langle\chi_{\mathrm{e}0}|\chi_{\mathrm{g}m}\right\rangle\right|^2\delta(E_{\mathrm{ZPL}}-E_{\mathrm{g}m}-\hbar\omega),
\end{equation}
where $\chi_{\mathrm{g(e)}m}$ is the $m$\textsuperscript{th} vibrational state of the ground (excited) electronic state and $\omega$ is the frequency of the transition.
$E_{\mathrm{g}m}$ represents the energy of the vibrational state $\chi_{\mathrm{g}m}$.
The true emission lineshape is determined by Fermi's golden rule, and is proportional to this overlap and the transition frequency 
as $L(\hbar\omega) \propto \omega^3 A(\hbar\omega)$.

\begin{figure}[t]
\includegraphics[width=\columnwidth]{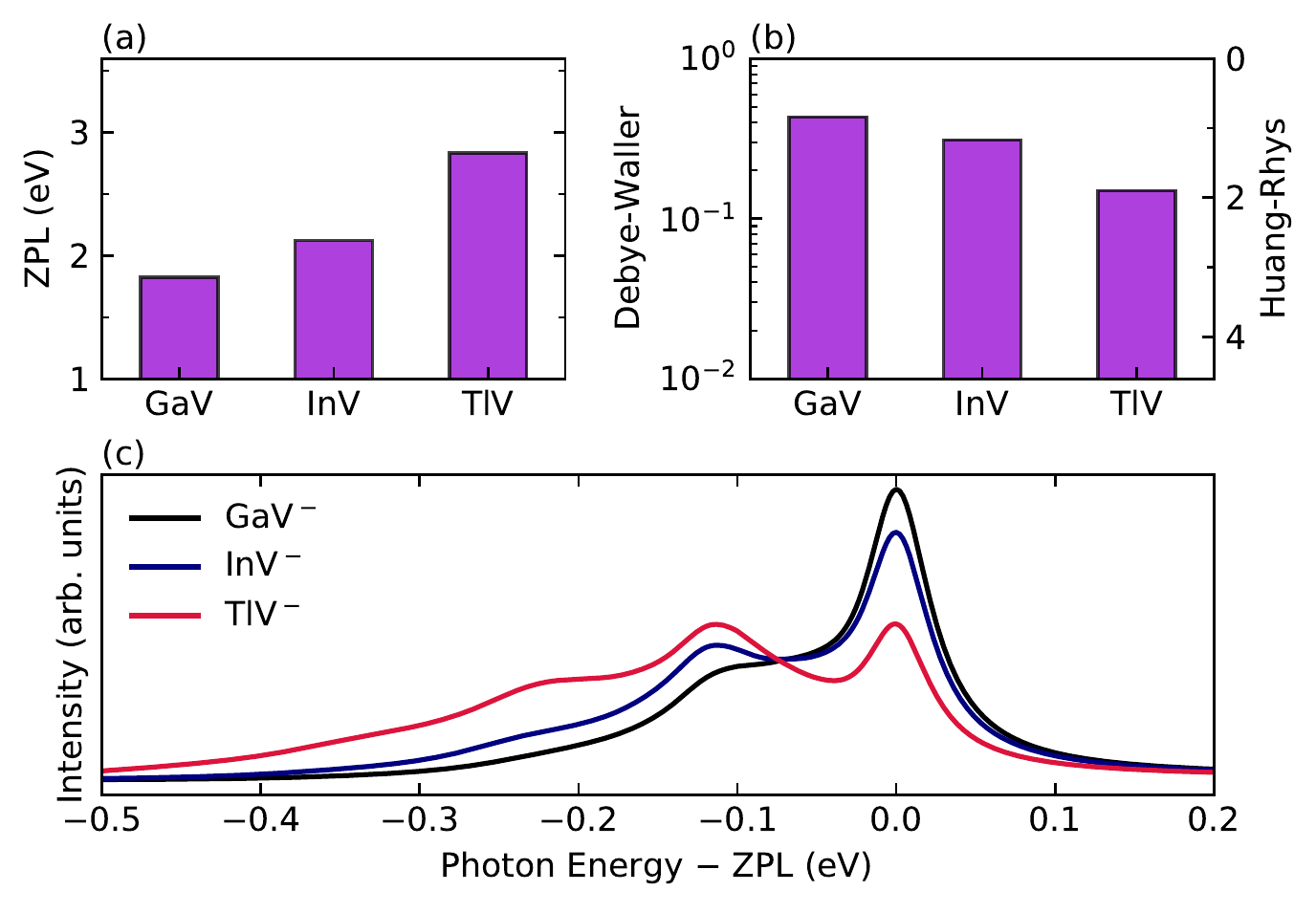}
\caption{\textbf{(a)} Predicted zero phonon line energy for each of the color centers found using $\Delta$SCF, including the pJT energy reduction in the excited state. \textbf{(b)} Debye-Waller factor (left) and Huang-Rhys factor (right) for each color center. The Debye-Waller factor indicates the fraction of photons emitted into the ZPL, while the Huang-Rhys factor indicates mean number of phonons emitted with each photon. \textbf{(c)} \emph{Ab initio} spectra for the $\dddd\rightarrow\dddd$ transition showing 
the predicted emission lineshape relative to the zero phonon line. We include linewidth broadening associated with  near room temperature conditions.
}
\label{fig:optical}
\end{figure}

The vibrational overlaps are evaluated using a generating function approach as described previously~\cite{Alkauskas2014}.
The generating function method assumes that the modes in the ground and excited state are identical, and there is only a difference in geometry between the two configurations.
This assumption is strongly violated when a Jahn-Teller distortion occurs in the excited state, since the symmetries between the two geometries are different.
Using the $\chh\rightarrow\dddd$ transition therefore underestimates the overlap of the ground vibrational states in the ground and excited state electronic manifolds.
Previous theoretical approaches for generating emission spectra for JT-unstable defects in diamond have neglected
the effect of the JT distortion by equally occupying the degenerate electronic orbitals using fractional occupations, effectively enforcing a high-symmetry configurations
(e.g., in the N\Vm~\cite{Alkauskas2014} and Si\Vm~\cite{londero_vibrational_2018}).
We use a similar approximation by studying the phonon coupling to the emission from the excited state with the correct orbital occupations, but restricted to a
\dddd\ geometry.
We note that this leads to a slightly lower Debye-Waller factor than the fractional occupation technique for the excited electronic states.

The lineshapes give information on what to expect from photoluminescence experiments. The Debye-Waller factor, 
which describes the overall emission efficiency into the ZPL,  is 
an important figure of merit for the performance of these defects in quantum protocols such as photon-mediated entanglement~\cite{Aharonovich:2016yu}. 
The resulting spectra, along with the predicted Huang-Rhys and Debye-Waller factors,
are given in Fig.~\ref{fig:optical}. The emission efficiency into the ZPL
is much higher than what is observed in N\Vm~\cite{Doherty2013}, although we find that the efficiency decreases for heavier defect elements. 

In conclusion, we predict the quantum optoelectronic properties of diamond group III vacancy centers.
Our calculations show that these color centers are stable in intrinsic diamond in a $S = 1$ spin state with a symmetry that makes their transitions insensitive to electric fields.
The combination of favorable thermodynamics, stable spin, and symmetry-protected optical properties makes the negatively charged group III vacancy centers ideal candidates for applications such as quantum networking and computing. The emerging ability to perform detailed calculations on spin and optical properties of nanoscopic quantum emitters, including corrections such as the large product Jahn-Teller effect, marks an important step towards \emph{ab initio} design and discovery of quantum materials. In particular, the predicted properties of the group III vacancy class of emitters would combine the desired attributes of high Debye-Waller factors with highly coherent spin and optical transitions at a range of wavelengths, promising a new class of quantum emitters for applications in quantum information science and technology. 

\section*{Acknowledgments}
We thank Dr. Matthew Trusheim (MIT-Harvard) for helpful discussions.
This work was supported by the DOE `Photonics at Thermodynamic Limits' Energy Frontier Research Center under grant number DE-SC0019140.
D. E. and P.N. are partially supported by the Army Research Office MURI (Ab-Initio Solid-State Quantum Materials)
grant number W911NF-18-1-0431. 
J.F. acknowledges fellowship support from the Deutsche Forschungsgemeinschaft (DFG) under Contract No. FL 997/1-1. P.N. is a Moore Inventor Fellow and a CIFAR Azrieli Global Scholar.

Calculations were performed using 
resources from the Department of Defense High Performance Computing
Modernization program. Additional calculations were performed 
using resources of the National Energy Research
Scientific Computing Center, a DOE Office of Science User
Facility, as well as resources at the Research Computing
Group at Harvard University.

 \end{document}